\documentclass[aps,prb,twocolumn,amsmath,amssymb,nofootinbib,superscriptaddress,floatfix,eqsecnum,]{revtex4}

\usepackage{amsmath}
\usepackage{amssymb}
\usepackage{amsthm}
\usepackage[pdftex]{color} 
\usepackage{graphicx}
\usepackage{dcolumn} 
\usepackage{bm} 
\usepackage{epic}{\usepackage[hypertex]{hyperref}
\usepackage{longtable}
\usepackage{ulem}   
\normalem           

\newcommand{\s}[1]{{\mathsf{#1}}} 
\newcommand{\bs}[1]{{\boldsymbol{#1}}}

\begin{document}

\title{
Time-reversal symmetric hierarchy of fractional incompressible liquids
      }

\author{Luiz Santos} 
\affiliation{
Department of Physics, 
Harvard University, 
17 Oxford St., 
Cambridge, MA 02138, USA
            } 

\author{Titus Neupert} 
\affiliation{
Condensed matter theory group, 
Paul Scherrer Institute, CH-5232 Villigen PSI,
Switzerland
            } 

\author{Shinsei Ryu} 
\affiliation{
Department of Physics, University of Illinois, 
1110 West Green St, Urbana IL 61801, USA
            } 

\author{Claudio Chamon} 
\affiliation{
Physics Department, 
Boston University, 
Boston, MA 02215, USA
            } 

\author{Christopher Mudry} 
\affiliation{
Condensed matter theory group, 
Paul Scherrer Institute, CH-5232 Villigen PSI,
Switzerland
            } 

\date{\today}

\begin{abstract}
We provide an effective description of fractional topological insulators
that include the fractional quantum spin Hall effect
by considering the time-reversal symmetric
pendant to the topological quantum field theories that encode
the Abelian fractional quantum Hall liquids.
We explain the hierarchical construction of such a theory and
establish for it a bulk-edge correspondence by
deriving the equivalent edge theory for chiral bosonic fields.
Further, we compute the Fermi-Bose correlation functions 
of the edge theory and provide representative ground state wave functions 
for systems described by the bulk theory.
\end{abstract}

\maketitle


\medskip
\section{
Introduction
        }
\label{sec: introduction}

Laughlin initiated the theoretical exploration of the
fractional quantum Hall effect (FQHE) by proposing wave functions for the
ground states of interacting electrons in the lowest Landau level at filling
fractions 
$\nu=1/(2m+1),\ m\in \mathbb{Z}$.%
~\cite{Laughlin83a}
The experimental observation of a plethora of fractional Hall plateaus 
at other filling fractions lead to the construction of a hierarchy 
of wave functions out of Laughlin's wave function,%
~\cite{Halperin83,Haldane83,Halperin84,Laughlin84,Girvin84,MacDonald85}
and the development of the composite fermion picture.%
~\cite{Jain89}
These approaches were later reconciled,
and unified by the effective description of the FQHE 
in terms of multi-component Chern-Simons theories
in $(2+1)$-dimensional space and time.%
~\cite{Read90,Wen90,Blok90,Frohlich91b,Wen92a,Wen92b,Frohlich94}
These topological effective theories for the hierarchy of the FQHE 
deliver a correspondence between the physics in the
two-dimensional bulk and the physics along one-dimensional boundaries
at which the two-dimensional sample terminates.%
~\cite{Laughlin81,Halperin82,Wen90b,Wen91a,Wen91b,Frohlich91a}

It is possible to double the Chern-Simons effective theory representing
the universal properties of the FQHE at some filling fraction
$\nu=1/(2m+1),\ m\in \mathbb{Z}$
so as to obtain a time-reversal symmetric theory.
This approach has been used 
to interpret a fully gaped superconductor as an example of a 
topological phase,%
~\cite{Wen89,Wen90a,Diamantini06}
and -- more generally -- to explore the universal properties of
interacting theories with an emergent local 
$\mathbb{Z}^{\ }_{2}$ gauge symmetry 
(see Refs.~\onlinecite{Read91,Wen91,Mudry94,Senthil00,Moessner01})
that signals the phenomenon of spin and charge separation.%
~\cite{Freedman04,Hansson04,kou05,kou08,Xu09}

A more urgent impetus for the construction
of effective time-reversal symmetric topological field theories
in $(2+1)$-dimensional space and time
arose with the theoretical prediction of
time-reversal symmetric topological band insulators,
shortly followed by their experimental discovery.%
~\cite{Kane05a,Kane05b,Bernevig06a,Bernevig06b,Konig07}
These band insulators
realize the counterparts to the integer quantum Hall effect
and their discovery suggests that a time-reversal symmetric 
counterpart to the FQHE might emerge from interacting itinerant
electrons in a crystalline environment.
 
From the outset, this endeavor follows a different line of logic 
than the FQHE, as it is not based on pre-existing experimental evidence.
Past experience with the FQHE has thus 
guided recent attempts to either construct
time-reversal symmetric edge theories 
or to construct time-reversal symmetric bulk wave functions supporting
local excitations carrying fractional quantum numbers.%
~\cite{Bernevig06a,Levin09,Cho11a,Cho11b,Qi11,Neupert11b,Goerbig11}
 
While numerical support for a time-reversal symmetric
topological phase of matter was given by Neupert et al.\
in their study of a lattice model for interacting itinerant electrons,%
~\cite{Neupert11b} 
a description in terms of an effective theory is
desirable to reveal the universal properties of such a phase.
In Ref.~\onlinecite{Neupert11b}, the universal properties such
as the topological degeneracies of the ground state manifold
were explored with the help of a family of edge theories.
In this paper, we are going to construct the corresponding bulk 
topological theory by generalizing the hierarchy of Abelian
FQHEs to the hierarchy of Abelian 
fractional quantum spin Hall effects (FQSHEs)
in Sec.%
~\ref{sec: TRS Abelian Chern-Simons quantum field theory}.
We will show in Sec.%
~\ref{sec: Bulk-edge correspondence}
the correspondence between the bulk theory and the edge theory
whose stability to the breaking of translation invariance
and residual spin-1/2 U(1) symmetry
was studied in Ref.~\onlinecite{Neupert11b}.
Finally, we shall generalize in Sec.%
~\ref{sec: Wave functions}
the wave functions supporting
the Abelian FQHE for a fractional filling of the lowest Landau level
to wave functions supporting an Abelian FQSHE.
These time-reversal symmetric wave functions are built from
the holomorphic and antiholomorphic 
single-particle wave functions belonging to the lowest Landau level
when the applied uniform magnetic field is pointing down or up,
respectively. For the reader who wants to skip
the derivations, we provide a detailed summary of our results
in Sec.~\ref{sec: Summary}.

\medskip
\section{
Time-reversal symmetric Abelian Chern-Simons quantum field theory
        }
\label{sec: TRS Abelian Chern-Simons quantum field theory}

Let us start by summarizing some of the results that we will derive in
this section. We shall construct a class of incompressible liquids,
each of which is the ground state of a time-reversal symmetric
(2+1)-dimensional Chern-Simons quantum field theory that depends on
$2N$ flavors of gauge fields
$a^{\ }_{i,\mu}(t,\bm{x})$, where $i=1,\dots,2N$ labels the flavors and
$\mu=0,1,2$ labels the space-time coordinates $x^{\mu}\equiv(t,\bm{x})$,
with the action
\begin{subequations}
\label{eq:intro-CS}
\begin{equation}
\begin{split}
\mathcal{S}:=
\int\mathrm{d}t\,\mathrm{d}^{2}\bs{x}\;
&
\epsilon_{\ }^{\mu\nu\rho} \,
\left(
-
\frac{1}{4\pi}
K^{\ }_{ij}\;
a^{\ }_{i,\mu}\;
\partial^{\ }_{\nu}\,
a^{\ }_{j,\rho}
\right.
\\&
\left.
+
\frac{e}{2\pi}\,Q^{\ }_{i}\,
A^{\ }_{\mu}\,
\partial^{\ }_{\nu}\,
a^{\ }_{i,\rho}
+
\frac{s}{2\pi}\,S^{\ }_{i}\,
B^{\ }_{\mu}\,
\partial^{\ }_{\nu}\,
a^{\ }_{i,\rho}
\right).
\end{split}
\label{eq:intro-CS a}
\end{equation}
Here, $K^{\ }_{ij}$ are elements of the symmetric and
invertible $2N\times2N$ integer matrix $K$. The integer-valued component
$Q^{\,}_{i}$ of the $2N$-dimensional vector $Q$ represents the $i$-th
electric charge in units of the electronic charge $e$, which couples
to the electromagnetic gauge potential $A_{\mu}(t,\bm{x})$. Similarly,
$S^{\,}_{i}$ is an integer-valued component of the $2N$-dimensional
vector $S$ that represents the $i$-th spin charge in units of $s$  
associated to the up or down spin projection 
along a spin-1/2 quantization axis,
which couples to the Abelian (spin) gauge potential
$B^{\,}_{\mu}(t,\bm{x})$. The operation of time reversal maps
$a^{\ }_{\s{i},\mu}(t,\bm{x})$ into $-g^{\mu\nu}\,
a^{\ }_{\s{i}+N,\nu}(-t,\bm{x})$ for $\s{i}=1,\cdots,N$ and vice versa.  
Here, $g^{\ }_{\mu\nu}:=\mathrm{diag}(+,-,-)\equiv g^{\mu\nu}$ 
is the Lorentz metric.  In
Eq.~(\ref{eq:intro-CS a}) $\bm{x}\in\Omega$, where
$\Omega\subset\mathbb{R}^{2}$ is a region of two-dimensional Euclidean
space, which for the discussion of the bulk theory we consider to have
no boundary, $\partial\,\Omega=\varnothing$. The domain of integration
$\mathbb{R}$ is unbounded in time $t$.
We will show that time-reversal symmetry imposes that the matrix
$K$ and the vectors $Q$ and $S$ are of the block form
\begin{equation}
K=\left(
\begin{matrix}
    \kappa&\;\Delta\\
    \Delta^{\!\mathsf{T}}&-\kappa
\end{matrix}
\right),
\quad
Q=\left(
\begin{matrix}
\varrho\\
\varrho
\end{matrix}
\right),
\quad
S=\left(
\begin{matrix}
\,\varrho\\
-\varrho
\end{matrix}
\right),
\label{eq:intro-K-matrix}
\end{equation}
\end{subequations}
with $\varrho$ an integer $N$-vector, while 
$\kappa=\kappa^{\!\mathsf{T}}$ and 
$\Delta=-\Delta^{\!\mathsf{T}}$
are symmetric and antisymmetric integer-valued $N\times N$ matrices,
respectively.

The doubled structure of the theory is even more evident if we express
it as a BF theory,~\cite{BFnote,Blau91,Blasi11} i.e., by defining 
\begin{subequations}
\begin{equation}
a^{(\pm)}_{\s{i},\mu}
:=
\frac{1}{2}
\left(
a^{\ }_{\s{i},\mu}
\pm
a^{\ }_{\s{i}+N,\mu}
\right),
\qquad \s{i}=1,\dots,N
\;,
\label{eq: def a +- a}
\end{equation}
for $\mu=0,1,2$.
This basis allows to re-express the effective action%
~(\ref{eq:intro-CS a}) as
\begin{equation}
\begin{split}
\mathcal{S}:=
\!\!\int\!\! \,\mathrm{d}t\, \mathrm{d}^{2}\bm{x}\; 
&\epsilon^{\mu\nu\rho} 
\left(
-
\frac{1}{\pi}
\varkappa^{\ }_{\s{ij}}\,
a^{(+)}_{\s{i},\mu}\,
\partial^{\ }_{\nu}\,
a^{(-)}_{\s{j},\rho}
\right.
\\
&
\left.
+ 
\frac{e}{\pi}\;
\rho^{\ }_{\s{i}}\,
A^{\ }_{\mu}\partial^{\ }_{\nu}\,
a^{(+)}_{\s{i},\rho}
+ 
\frac{s}{\pi}\;
\rho^{\ }_{\s{i}}\,
B^{\ }_{\mu}\partial^{\ }_{\nu}\,
a^{(-)}_{\s{i},\rho}
\right).
\label{eq:intro-CS-BF}
\end{split}
\end{equation}
In this representation, the indices in sans serif fonts
$\s{i},\s{j}$ run from
$1$ to $N$. The coupling between the pair of gauge fields
$a^{(+)}$ and $a^{(-)}$ is off-diagonal in the BF labels $\pm$. 
This is a consequence of time-reversal
symmetry, which is implemented by
\begin{equation}
a^{(\pm)}_{\mu}(t,\bs{x})
\stackrel{\mathcal{T}}{\rightarrow}
\mp g^{\mu\nu}\, 
a^{(\pm)}_{\nu}(-t,\bs{x})
\;,
\label{eq: def TR bf}
\end{equation}
that leaves the action%
~(\ref{eq:intro-CS-BF}) invariant. In this representation, the 
electromagnetic gauge potential $A$ couples to the
$+$-species only, while the spin gauge potential $B$ couples to the 
$-$-species only. The $N\times N$ integer-valued matrix $\varkappa$ 
in the BF representation is related to the 
block matrices $\kappa$ and $\Delta$ contained in $K$ from 
Eq.~(\ref{eq:intro-K-matrix})
through
\begin{equation}
\varkappa=
\kappa
-
\Delta.
\end{equation}
\end{subequations}

The degeneracy of the ground state is obtained for either
description, 
i.e., the one in terms of the flavors 
$a^{\ }_{i}$ with $i=1,\cdots,2N$ 
or the one in terms of the flavors 
$a^{(\pm)}_{\s{i}}$ with $\s{i}=1,\cdots,N$, 
from
\begin{eqnarray}
\label{eq:intro-degeneracy}
\mathcal{N}^{\ }_{\mathrm{GS}}&=&
\left|
\det
\begin{pmatrix}
0
&
\varkappa
\\
\varkappa^{\!\mathsf{T}}
&
0
\end{pmatrix}
\right|
=\left(\det \varkappa\right)^{2}
.
\end{eqnarray}

If the underlying microscopic theory describes fermions with
a residual spin-1/2 U(1) (easy plane $XY$)
symmetry, it is then meaningful to define the 
quantized spin Hall resistance
\begin{subequations}
\label{eq: QSHE}
\begin{equation}
\sigma^{\ }_{\mathrm{sH}}:=
\frac{e}{2\pi}\times\nu^{\ }_{\mathrm{s}}.
\label{eq: QSHE a}
\end{equation}
The filling fraction $\nu^{\ }_{\mathrm{s}}$
is here defined so that it is unity for the integer
quantum spin Hall effect 
and therefore given by
\begin{equation}
\begin{split}
\nu^{\ }_{\mathrm{s}}:=&\,
\frac{1}{2}
Q^{\s{T}}\, K^{-1}\,S
\\
=&\,
\varrho^{\s{T}}\,\varkappa^{-1}\,\varrho.
\end{split}
\label{eq: QSHE c}
\end{equation}
\end{subequations}

We now turn to the hierarchical construction of the states described by
this quantum field theory. 
As a warm-up, we begin by reviewing how a one-component
Chern-Simons quantum field theory in $(2+1)$-dimensional space and
time is related to the quantum Hall effect. We then construct
recursively the multi-component Chern-Simons quantum field theory in such
a way that it respects time-reversal symmetry.

\subsection{
Brief review of the one-component Chern-Simons theory
           }
\label{subsec: Hall conductance from a one-component Chern-Simons theory}

We start from the Lagrangian density 
\begin{subequations}
\label{eq: def FQHE 1-component}
\begin{equation}
\mathcal{L}^{\ }_{\mathrm{CS}}:=
-
\frac{p}{4\pi}\,
\epsilon^{\mu\nu\lambda}\,
a^{\ }_{\mu}\,\partial^{\ }_{\nu}\,a^{\ }_{\lambda}
+
\frac{e}{2\pi}\,
\epsilon^{\mu\nu\lambda}\,
A^{\ }_{\mu}\,
\partial^{\ }_{\nu}\,a^{\ }_{\lambda}	
\label{eq: def FQHE 1-component a}
\end{equation}
in ($2+1$)-dimensional space and time with the action
\begin{equation}
\mathcal{S}^{\ }_{\mathrm{CS}}:= 
\int\limits_{\mathbb{R}}\mathrm{d}t\,
\int\limits_{\Omega}\mathrm{d}^{2}\bm{x}\, 
\mathcal{L}^{\ }_{\mathrm{CS}}
\label{eq: def FQHE 1-component b}
\end{equation}
and partition function
\begin{equation}
Z^{\ }_{\mathrm{CS}}[A]:= 
\int\limits
\mathcal{D}[a]\,
e^{\frac{\mathrm{i}}{\hbar}\mathcal{S}^{\ }_{\mathrm{CS}}}.
\end{equation}
\end{subequations}
The dimensionless integer $p$ is positive.
The electromagnetic coupling (electric charge)
$e$ is dimensionfull. It measures the strength of the
interaction between an external electromagnetic gauge field 
$A$ with the components
$
A^{\mu}\equiv(A^{0},\bm{A}) 
$
and a dynamical gauge field 
$a$ 
with the components
$
a^{\mu}\equiv(a^{0},\bm{a}) 
$.
The symbol $\mathcal{D}[a]$ represents the measure of all gauge orbits 
stemming from the Abelian group U(1).

The operation $\mathcal{T}$ for reversal of time
is defined by
\begin{subequations}
\begin{equation}
a^{\ }_{\mu}(t,\bs{x})
\stackrel{\mathcal{T}}{\rightarrow}\,
+
g^{\mu\nu}\,
a^{\ }_{\nu}(-t,\bs{x}),
\end{equation}
\begin{equation}
A^{\ }_{\mu}(t,\bs{x})
\stackrel{\mathcal{T}}{\rightarrow}\,
+
g^{\mu\nu}\,
A^{\ }_{\nu}(-t,\bs{x}),
\end{equation}
\end{subequations}
for $\mu=0,1,2$.
We also posit that $\mathcal{T}$ is
an anti-unitary linear transformation. 
If so, one verifies that
$\mathcal{L}^{\ }_{\mathrm{CS}}$
is odd under reversal of time.

Define the 
electromagnetic
current to be the 3-vector
\begin{subequations}
\label{eq: def Jmu for one CS field}
\begin{equation}
\begin{split}
J^{\mu}_{\mathrm{CS}}:=&\,
\frac{1}{\hbar}
\frac{
\delta \mathcal{S}^{\ }_{\mathrm{CS}}
     }
     {
\delta A^{\ }_{\mu}
     }
\\
=&\,
\frac{e}{2\pi\hbar}\,
\epsilon^{\mu\nu\lambda}\,
\partial^{\ }_{\nu}\,a^{\ }_{\lambda}
\end{split}
\end{equation}
for $\mu=0,1,2$.
Because the Levi-Civita tensor with the component 
$\epsilon^{012}\equiv1$
is fully antisymmetric,
this current is conserved,
\begin{equation}
\partial^{\ }_{\mu}J^{\mu}_{\mathrm{CS}}=0.
\end{equation}
\end{subequations}
Now, the equations of motions 
\begin{equation}
\begin{split}
0=&\,
\frac{
\delta \mathcal{S}^{\ }_{\mathrm{CS}}
     }
     {
\delta a^{\ }_{\mu}
     }
=
-
\frac{p}{2\pi}
\epsilon^{\mu\nu\lambda}\,
\partial^{\ }_{\nu}\,
a^{\ }_{\lambda}
+
\frac{e}{2\pi}
\epsilon^{\mu\nu\lambda}\,
\partial^{\ }_{\nu}\,
A^{\ }_{\lambda}
\end{split}
\end{equation}
can be used in conjunction with Eq.~(\ref{eq: def Jmu for one CS field})
to yield the conserved electromagnetic current
\begin{equation}
\begin{split}
J^{\mu}_{\mathrm{CS}}=&\,
\frac{1}{p}\,\frac{e^{2}}{h}\,
\epsilon^{\mu\nu\lambda}\,
\partial^{\ }_{\nu}\,
A^{\ }_{\lambda}
\end{split}
\end{equation}
which allows us to identify the filling fraction $\nu=p^{-1}$ in this
simple example, so that the quantum Hall conductance is given by $\sigma_{\mathrm H}=\nu\, \frac{e^{2}}{h}$.
From now on, we adopt units in which $\hbar = 1$.

\subsection{
One-component BF theory
           }
\label{subsec: Double Chern-Simons theory with time-reversal symmetry}

We start from the Lagrangian density in 
($2+1$)-dimensional space and time
\begin{subequations}
\label{eq: def TRS CS level 0}
\begin{equation}
\begin{split}
\mathcal{L}^{\mathrm{TRS}}_{\mathrm{BF}}:=&\,
-
\frac{p}{\pi}\,
\epsilon^{\mu\nu\lambda}\,
a^{(+)}_{\mu}\,\partial^{\ }_{\nu}\,a^{(-)}_{\lambda}
\\
&\,
+
\frac{e}{\pi}\,
\epsilon^{\mu\nu\lambda}\,
A^{\ }_{\mu}\,
\partial^{\ }_{\nu}\,a^{(+)}_{\lambda}
+
\frac{s}{\pi}\,
\epsilon^{\mu\nu\lambda}\,
B^{\ }_{\mu}\,
\partial^{\ }_{\nu}\,a^{(-)}_{\lambda}
\end{split}
\label{eq: def TRS CS level 0 a}
\end{equation}
with the action
\begin{equation}
\mathcal{S}^{\mathrm{TRS}}_{\mathrm{BF}}:= 
\int\limits_{\mathbb{R}}\mathrm{d}t\,
\int\limits_{\Omega}\mathrm{d}^{2}\bm{x}\, 
\mathcal{L}^{\mathrm{TRS}}_{\mathrm{BF}}
\end{equation}
and partition function
\begin{equation}
Z^{\mathrm{TRS}}_{\mathrm{BF}}[A,B]:= 
\int\mathcal{D}[a^{(+)},a^{(-)}]\,
e^{\mathrm{i}\mathcal{S}^{\mathrm{TRS}}_{\mathrm{BF}}}.
\label{eq: def TRS CS level 0 c}
\end{equation}
\end{subequations}
Equation~(\ref{eq: def TRS CS level 0})
is a BF theory made of two copies of the Chern-Simons theory%
~(\ref{eq: def FQHE 1-component})
with the specificity that the integer $p$ enters
with opposite signs in the two copies.
We have also 
introduced
two external gauge fields $A$ and $B$
with the couplings $e$ and $s$, respectively.
For the gauge field $A$, $e$ will be
interpreted as a total U(1) charge. For the gauge field $B$,
$s$
will be interpreted as a relative U(1) charge.
If the underlying microscopic model is built from itinerant
electrons,
the gauge field $A$ is the U(1) electromagnetic
gauge field that couples to the conserved electric charge whereas
the gauge field $B$ is the U(1) gauge field that couples to the 
conserved projection along some quantization axis 
of the electronic spin,
i.e., $s=1/2$.

This theory is invariant under the operation of time
reversal defined by the anti-linear extension of
\begin{subequations}
\label{eq: def TR for two component}
\begin{equation}
a^{(\pm)}_{\mu}(t,\bs{x})
\stackrel{\mathcal{T}}{\rightarrow}\,
\mp
g^{\mu\nu}\,
a^{(\pm)}_{\nu}(-t,\bs{x})\equiv
\mp a^{(\pm)\mu}(\tilde{t},\tilde{\bs{x}}),
\end{equation}
\begin{equation}
A^{\ }_{\mu}(t,\bs{x})
\stackrel{\mathcal{T}}{\rightarrow}\,
+
g^{\mu\nu}\,
A^{\ }_{\nu}(-t,\bs{x})\equiv
+
A^{\mu}(\tilde{t},\tilde{\bs{x}}),
\label{eq: def TR for two component c}
\end{equation}
\begin{equation}
B^{\ }_{\mu}(t,\bs{x})
\stackrel{\mathcal{T}}{\rightarrow}\,
-
g^{\mu\nu}\,
B^{\ }_{\nu}(-t,\bs{x})\equiv
-
B^{\mu}(\tilde{t},\tilde{\bs{x}}),
\label{eq: def TR for two component d}
\end{equation}
\end{subequations}
for $\mu=0,1,2$. 
The component $A^{0}$ of the external electromagnetic gauge field $A$
is unchanged whereas its vector component $\bm{A}$ 
is reversed under reversal of time, just as the vector components of $a^{(-)}$.
This behavior is reversed for the components of the
external gauge field $B$ that couples to the conserved U(1) spin current and the gauge field $a^{(+)}$.

Since this theory is equivalent to two independent copies of
the Chern-Simons theory%
~(\ref{eq: def FQHE 1-component}),
there are two independent conserved currents of the form%
~(\ref{eq: def Jmu for one CS field}),
\begin{equation}
J^{\mu}_{\pm}:=
\frac{e}{\pi}
\epsilon^{\mu\nu\lambda}
\partial^{\vphantom{(\pm)}}_{\nu}
a^{(\pm)}_{\lambda},
\end{equation}
for $\mu=0,1,2$.
Their transformation laws under reversal of time are
\begin{equation}
J^{\mu}_{\pm}(x)\stackrel{\mathcal{T}}{\rightarrow}\,
\pm
g^{\ }_{\mu\nu}
J^{\nu}_{\pm}(\tilde{x}),
\end{equation}
for $\mu=0,1,2$.
If the microscopic model is made of itinerant electrons, we can thus interpret $J^{\mu}_{+}$ as 
the charge current and, if the model has a 
residual U(1) rotation symmetry of the electronic spin, $J^{\mu}_{-}$ represents the conserved spin current.
The equations of motions 
\begin{subequations}
\begin{equation}
0=
\frac{
\delta \mathcal{S}^{\mathrm{TRS}}_{\mathrm{BF}}
     }
     {
\delta a^{(\pm)}_{\mu}
     }
\end{equation}
for the dynamical compact gauge fields 
$a^{(-)}$ and $a^{(+)}$, respectively, deliver the relations
\begin{equation}
\epsilon^{\mu\nu\lambda}\,
\partial^{\ }_{\nu}\,
a^{(+)}_{\lambda}= 
\frac{s}{p}\,
\epsilon^{\mu\nu\lambda}\,
\partial^{\ }_{\nu}\,
B^{\ }_{\lambda}	
\end{equation}
and
\begin{equation}
\epsilon^{\mu\nu\lambda}\,
\partial^{\ }_{\nu}\,
a^{(-)}_{\lambda}=
\frac{e}{p}\,
\epsilon^{\mu\nu\lambda}\,
\partial^{\ }_{\nu}\,
A^{\ }_{\lambda},
\end{equation}
\end{subequations}
for $\mu=0,1,2$, respectively. We conclude that,
on the one hand, 
the charge current obeys the Hall response 
\begin{subequations}
\begin{equation}
J^{\mu}_{+}
=
2 s \times
\frac{e}{2\pi p}
\epsilon^{\mu\nu\lambda}\,
\partial^{\ }_{\nu}\,
B^{\ }_{\lambda},
\end{equation}
with $\mu=0,1,2$ 
while, on the other hand, 
the spin current obeys the Hall response 
\begin{equation}
J^{\mu}_{-}
=
2e \times\frac{e}{2\pi p}\,
\epsilon^{\mu\nu\lambda}\,
\partial^{\ }_{\nu}\,
A^{\ }_{\lambda},
\end{equation}
\end{subequations}
with $\mu=0,1,2$.

\subsection{
Time-reversal symmetric hierarchy
           }
\label{subsec: n-step hierarchy construction}

The generic structure of the hierarchical construction is the following.
Let $n>0$ be any positive integer. Define at the level 
$n$ of the hierarchy the quantum field theory 
with the partition function
\begin{subequations}
\label{eq: def level n}
\begin{equation}
\begin{split}
Z^{\mathrm{TRS}}_{n}[A,B]:=&\,
\int\mathcal{D}
\left[
a^{(+)}_{1},\cdots,a^{(+)}_{n},
a^{(-)}_{1},\cdots,a^{(-)}_{n}
\right]\\
&\times
e^{\mathrm{i}\mathcal{S}^{\mathrm{TRS}}_{n}},
\label{eq: def level n a}
\end{split}
\end{equation}
where the action is
\begin{equation}
\mathcal{S}^{\mathrm{TRS}}_{n}:=
\int\limits_{\mathbb{R}}\mathrm{d}t\,
\int\limits_{\Omega}\mathrm{d}^{2}\bm{x}\, 
\mathcal{L}^{\mathrm{TRS}}_{n}
\label{eq: def level n b}
\end{equation}
and the Lagrangian density is
\begin{equation}
\begin{split}
\mathcal{L}^{\mathrm{TRS}}_{n}:=&\,
-\sum_{\s{i},\s{j}=1}^{n}\,
\frac{1}{\pi}\,
\varkappa^{(n)}_{\s{i}\s{j}}\,
\epsilon^{\mu\nu\lambda}\,
a^{(+)}_{\s{i},\mu}\,
\partial_{\nu}\,
a^{(-)}_{\s{j},\lambda}
\\
&\,
+
\sum_{\s{i}=1}^{n}\,
\frac{e}{\pi}\,
\varrho^{(n)}_{\s{i}}\,
\epsilon^{\mu\nu\lambda}\,
A^{\ }_{\mu}\,
\partial_{\nu}\,
a^{(+)}_{\s{i},\lambda}
\\
&\,
+
\sum_{\s{i}=1}^{n}\,
\frac{s}{\pi}\,
\varrho^{(n)}_{\s{i}}\,
\epsilon^{\mu\nu\lambda}\,
B^{\ }_{\mu}\,
\partial_{\nu}\,
a^{(-)}_{\s{i},\lambda}
.
\end{split}
\label{eq: def level n c}
\end{equation}
\end{subequations}
Here, the dynamical gauge fields $a^{(\pm)}$ are the
$n$-tuplet with the components
\begin{subequations}
\begin{equation}
\left(a^{(\pm)}_{\s{i}}\right)\equiv
\left(a^{(\pm)}_{1},\cdots,a^{(\pm)}_{n}\right)^{\s{T}}.
\label{eq: def 2n tuplet}
\end{equation}
Moreover, the $n\times n$ matrix $\varkappa^{(n)}$
is invertible and has, by assumption, integer-valued matrix elements.
The charge vector $\varrho^{(n)}$ has the integer-valued components 
\begin{equation}
\varrho^{(n)}=
(1,0,\cdots,0)^{\s{T}}
\in\mathbb{Z}^{n}.
\end{equation}
Finally, the compatibility condition
\begin{equation}
(-)^{\varkappa^{(n)}_{\s{i}\s{i}}}=
(-)^{\varrho^{(n)}_{\s{i}}}
\end{equation}
for $\s{i}=1,\cdots,n$ is also assumed.
\end{subequations}

The operation of time reversal is the rule
\begin{subequations}
\label{eq: def TR level n}
\begin{equation}
x^{\mu}
\stackrel{\mathcal{T}}{\rightarrow}
\tilde{x}^{\mu}:=
-g^{\ }_{\mu\nu}\, 
x^{\nu}
\end{equation}
together with the anti-linear extension of the rules
\begin{equation}
a^{(\pm)\mu}_{\s{i}}(x)
\stackrel{\mathcal{T}}{\rightarrow}
\mp g^{\ }_{\mu\nu}\, 
a^{(\pm)\nu}_{\s{i}}(\tilde{x}),
\end{equation}
\end{subequations}
for $\mu=0,1,2$ and $\s{i}=1,\cdots,n$
that leaves the Lagrangian density%
~(\ref{eq: def level n c}) 
invariant.

The level $n+1$ of the hierarchical construction
posits the existence of
the pair of quasiparticle 3-currents 
$j^{\ }_{\pm,n+1}$ 
that are conserved, i.e.,
\begin{equation}
\partial^{\ }_{\mu}\,j^{\mu}_{\pm,n+1}=0.
\label{eq: flux attachment a}
\end{equation} 
It also posits the existence of
some \textit{even} integer $p^{\ }_{n+1}$ and $2n$ integers 
$l^{(+)}_{\s{i}},\ l^{(-)}_{\s{i}}$ with $\s{i}=1,\cdots,n$
such that the constraints
\begin{equation}
j^{\mu}_{\pm,n+1}=
\frac{\epsilon^{\mu\nu\lambda}}{\pi p^{\ }_{n+1}}\,
\sum_{\s{i}=1}^{n}
l^{(\pm)}_{\s{i}}\,
\partial^{\ }_{\nu}\,
a^{(\pm)}_{\s{i},\lambda}
\label{eq: flux attachment b}
\end{equation}
for $\mu=0,1,2$
hold. The constraint%
~(\ref{eq: flux attachment b})
means that any pair of flux quanta, 
arising when
$a^{(+)}_{\mathsf{i}}$
and 
$a^{(-)}_{\mathsf{i}}$
each support a vortex,
creates a quasi-particle with charge
$2\, l^{(+)}_{\s{i}}/p^{\ }_{n+1}$
and spin
$2\, l^{(-)}_{\s{i}}/p^{\ }_{n+1}$
for $\s{i}=1,\cdots,n$.

This construction can be achieved from the partition function
\begin{subequations}
\begin{equation}
\begin{split}
Z^{\mathrm{TRS}}_{n+1}[A,B]:=&\,
\int\mathcal{D}
\left[
a^{(+)}_{1},\cdots,a^{(+)}_{n+1},
a^{(-)}_{1},\cdots,a^{(-)}_{n+1}
\right]\\
&\times
e^{\mathrm{i}\mathcal{S}^{\mathrm{TRS}}_{n+1}},
\end{split}
\end{equation}
with the action
\begin{equation}
\mathcal{S}^{\mathrm{TRS}}_{n+1}:=
\int\limits_{\mathbb{R}}\mathrm{d}t\,
\int\limits_{\Omega}\mathrm{d}^{2}\bm{x}\, 
\mathcal{L}^{\ }_{n+1}
\end{equation}
and Lagrangian density
\begin{equation}
\label{Ln+1}
\begin{split}
\mathcal{L}^{\mathrm{TRS}}_{n+1}:=&\,
\mathcal{L}^{\mathrm{TRS}}_{n}
\\
&\,
-
\frac{p^{\ }_{n+1}}{\pi}\,
\epsilon^{\mu\nu\lambda}\,
a^{(+)}_{n+1,\mu}\,
\partial^{\ }_{\nu}\,
a^{(-)}_{n+1,\lambda}
\\
&\,
+
\frac{1}{\pi}\,
\epsilon^{\mu\nu\lambda}\,
\sum_{\mathsf{i}=1}^{n}
l^{(+)}_{\mathsf{i}}\,
a^{(+)}_{\mathsf{i},\mu}\,
\partial^{\ }_{\nu}\,
a^{(-)}_{n+1,\lambda}
\\
&\,
+
\frac{1}{\pi}\,
\epsilon^{\mu\nu\lambda}\,
\sum_{\mathsf{i}=1}^{n}
l^{(-)}_{\mathsf{i}}\,
a^{(-)}_{\mathsf{i},\mu}\,
\partial^{\ }_{\nu}\,
a^{(+)}_{n+1,\lambda}.
\end{split}
\end{equation}
\end{subequations}
Indeed,
we can then \textit{define} the conserved quasiparticle currents of type
$n$ to be
\begin{equation}
j^{\mu}_{\pm,n+1}:=
\frac{1}{\pi}\,
\epsilon^{\mu\nu\lambda}\,
\partial^{\ }_{\nu}\,                 
a^{(\pm)}_{n+1,\lambda}
\end{equation}
for $\mu=0,1,2$ and use the equations of motion 
\begin{equation}
\begin{split}
&
0=
\frac{
\delta\mathcal{S}^{\mathrm{TRS}}_{n+1}
     }
     {
\delta a^{(\mp)}_{n+1,\mu}
     }
\Longleftrightarrow
\\
&
\frac{p^{\ }_{n+1}}{\pi}
\epsilon^{\mu\nu\lambda}
\partial^{\ }_{\nu}
a^{(\pm)}_{n+1,\lambda}
=
\frac{\epsilon^{\mu\nu\lambda}}{\pi}\,
\sum_{\s{i}=1}^{n}
l^{(\pm)}_{\s{i}}\,
\partial^{\ }_{\nu}\,
a^{(\pm)}_{\s{i},\lambda}
\end{split}
\end{equation}
obeyed by the dynamical gauge fields 
$a^{(\pm)}_{n+1,\mu}$
to establish that they indeed obey the constraints imposed in 
Eq.~\eqref{eq: flux attachment b}.

Observe that if we introduce the two $(n+1)$-tuplets $a^{(\pm)}$ given by
\begin{subequations}
\label{eq: step n+1 rep K and Q}
\begin{equation}
\left(a^{(\pm)}_{\s{i}}\right)^{\s{T}}\equiv 
\left(a^{(\pm)}_{1},\cdots,a^{(\pm)}_{n+1}\right)^{\s{T}}
\label{eq: final def Ln+1 a}
\end{equation}
of dynamical gauge fields, then the Lagrangian 
$\mathcal{L}^{\mathrm{TRS}}_{n+1}$ defined in Eq.~\eqref{Ln+1}
takes the same form as $\mathcal{L}^{\mathrm{TRS}}_{n}$ defined 
in Eq.~\eqref{eq: def level n c}
after the substitution $n\to n+1$.
The $(n+1)\times (n+1)$ matrix $\varkappa^{(n+1)}$ is then given by
\begin{equation}
\varkappa^{(n+1)}= 
\begin{pmatrix}
\varkappa^{(n)}
& 
-l^{(+)}
\\
-l^{(-)\s{T}}
& 
p^{\ }_{n+1}
\end{pmatrix}.
\end{equation}
The $(n+1)$-component charge vector 
$\varrho^{(n+1)}$ is given by
\begin{equation}
\varrho^{(n+1)}=
(1,0,\cdots,0)^{\s{T}}\in\mathbb{Z}^{n+1},
\end{equation}
thus imposing a vanishing coupling of the external gauge fields $A$ and $B$ to 
$a^{(\pm)}_{n+1}$.
The compatibility condition
\begin{equation}
(-)^{\varkappa^{(n)}_{\s{ii}}}=
(-)^{\varrho^{(n)}_{\s{i}}}
\end{equation}
for $\s{i}=1,\cdots,n+1$ holds if and only if 
the integer $p^{\ }_{n+1}$ is even.
\end{subequations}

The representation%
~(\ref{eq: step n+1 rep K and Q})
is called the hierarchical representation.

The operation of time reversal obtained from
Eq.~(\ref{eq: def TR level n})
by allowing $\s{i}$ to run from 1 up to $n+1$ leaves
the Lagrangian of level $n+1$ invariant.
Therefore, we have constructed 
a hierarchical time-reversal symmetric BF theory.

\subsection{
Equivalent representations
        }
\label{subsec: Equivalent representations}

We define an equivalence class
on all the actions of the form%
~(\ref{eq:intro-CS-BF})
when there exists a linear transformation 
$W$
with integer valued coefficients and unit determinant
\begin{subequations}
\label{eq: def equivalence classes}
such that
\begin{equation}
\varkappa=
W^{\s{T}}\,
\varkappa'\,
W
\end{equation}
and
\begin{equation}
\varrho=
W^{\s{T}}\,
\varrho',
\end{equation}
\end{subequations}
between any two given pairs
$(\varkappa,\varrho)$ and $(\varkappa',\varrho')$
within an equivalence class.

\textit{Example 1:}
The lower-triangular transformation
\begin{subequations}
\label{eq: def trsf from hierarchy to symmetric basis}
\begin{equation}
W^{\mathsf{T}}:=
\begin{pmatrix}
1
&
0
&
\cdots
&
0
\\
1
&
-1
&
\cdots
&
0
\\
\vdots
&
\vdots
&
\cdots
&
\vdots
\\
1
&
0
&
\cdots
&
-1
\end{pmatrix}
\end{equation}
relates the hierarchical basis characterized 
by the charge vectors
\begin{equation}
\varrho=
(1,0,\cdots,0)^{\mathsf{T}}
\end{equation}
to the so-called symmetric basis
characterized by the charge vector
\begin{equation}
\varrho=
(1,1,\cdots,1)^{\mathsf{T}}.
\end{equation}
\end{subequations}

\textit{Example 2:}
The block-diagonal transformation
\begin{equation}
W^{\mathsf{T}}:=
\begin{pmatrix}
\openone^{\ }_{m-1}&      &                     &      &                   \\
                   &     0&                     &    -1&                   \\
                   &      &\openone^{\ }_{n-1-m}&      &                   \\
                   &    +1&                     &     0&                   \\
                   &      &                     &      &\openone^{\ }_{N-n}\\
\end{pmatrix}
\label{eq: example 2 for W}
\end{equation}
with $1\leq m<n\leq N$ that interchanges
$\varkappa^{\ }_{mm}$ with $\varkappa^{\ }_{nn}$,
$\varkappa^{\ }_{mn}$ with $-\varkappa^{\ }_{nm}$,
while it substitutes
$-\varrho^{\ }_{n}$ for $\varrho^{\ }_{m}$
and
$+\varrho^{\ }_{m}$ for $\varrho^{\ }_{n}$.

\medskip
\section{
Edge theory
        }
\label{sec: Bulk-edge correspondence}

In this Section, we study the quantum field theory for 
$2N$ Abelian Chern-Simons fields as defined in%
~(\ref{eq:intro-CS a}) or, equivalently,~(\ref{eq:intro-CS-BF}) 
in a system with a boundary by following a strategy pioneered in Refs.%
~\onlinecite{Elitzur89} and \onlinecite{Wen95}. 
However, before relaxing the condition 
$\partial\,\Omega=\varnothing$,
we decompose the action%
~(\ref{eq:intro-CS a}) of the bulk theory
into
\begin{subequations}
\label{eq:edge-CS}
\begin{equation}
\mathcal{S}:= 
\mathcal{S}^{\ }_{K} 
+ 
\mathcal{S}^{\ }_{Q} 
+ 
\mathcal{S}^{\ }_{S},	
\label{eq:edge-CS a}
\end{equation}
\begin{equation}
\mathcal{S}^{\ }_{K}:=
-\frac{1}{4\pi}
\int\limits_{\mathbb{R}}\mathrm{d}t\,
\int\limits_{\Omega}\mathrm{d}^{2}\bm{x}\,
K^{\ }_{ij}\, 
\epsilon_{\ }^{\mu\nu\rho}\,
a^{\ }_{i,\mu}\,
\partial^{\ }_{\nu}\,
a^{\ }_{j,\rho},
\label{eq:edge-CS b}
\end{equation}
\begin{equation}
\mathcal{S}^{\ }_{Q}:=
+
\int\limits_{\mathbb{R}}\mathrm{d}t\,
\int\limits_{\Omega}\mathrm{d}^{2}\bm{x}\,
\frac{e}{2\pi}\,Q^{\ }_{i}\,
\epsilon^{\mu\nu\rho}\,
a^{\ }_{i,\mu}\,
\partial^{\ }_{\nu}\,
A^{\ }_{\rho},
\label{eq:edge-CS c}	
\end{equation}
\begin{equation}
\mathcal{S}^{\ }_{S}:=
+
\int\limits_{\mathbb{R}}\mathrm{d}t\,
\int\limits_{\Omega}\mathrm{d}^{2}\bm{x}\,
\frac{s}{2\pi}\,S^{\ }_{i}\,
\epsilon_{\ }^{\mu\nu\rho}\,
a^{\ }_{i,\mu}\,
\partial^{\ }_{\nu}\,
B^{\ }_{\rho}.
\label{eq:edge-CS d}
\end{equation}
\end{subequations}
Notice that we have performed a partial integration in 
Eq.~\eqref{eq:edge-CS c}
and Eq.~\eqref{eq:edge-CS d} as compared to Eq.~\eqref{eq:intro-CS a}, 
so that the gauge fields $A$ and $B$ enter Eq.~\eqref{eq:edge-CS} 
in an explicitly gauge invariant form.
In contrast, we are going to make a gauge choice for the fields 
$a^{\ }_{i}$ with $i=1,\cdots,2N$ 
to derive the gauge-invariant effective theory of the edge, once
we have relaxed the condition $\partial\,\Omega=\varnothing$.

Let us choose $\Omega$ to be the upper-half plane of 
$\mathbb{R}^{2}$, i.e.,
\begin{equation}
\Omega:=
\left\{\left.(x,y)\in\mathbb{R}^{2}\right|
y\geq0
\right\}
\label{eq: def Omega}
\end{equation}
for notational simplicity but without loss of generality.
Observe that under the $2N$ independent Abelian gauge transformations
of the dynamical Chern-Simons fields 
\begin{subequations}
\label{eq: delta S due  a to a + da}
\begin{equation}
a^{\ }_{i,\mu}\to
a^{\ }_{i,\mu}
+
\partial^{\ }_{\mu} 
\chi^{\ }_{i} 
\label{eq: delta S due  a to a + da a}
\end{equation}
for $\mu=0,1,2$
where $\chi^{\ }_{i}$ 
with $i=1,\cdots,2N$ are real-valued and smooth,
the action $\mathcal{S}$ defined in Eq.~(\ref{eq:edge-CS}) 
obeys the transformation law
\begin{equation}
\mathcal{S}\to
\mathcal{S}
+
\delta\mathcal{S}
\label{eq: delta S due  a to a + da b}
\end{equation}
with
\begin{equation}
\delta\mathcal{S}=
\int\limits_{-\infty}^{+\infty}\mathrm{d}\,t
\int\limits_{-\infty}^{+\infty}\mathrm{d}\,x\,
\left(\chi^{\ }_i\,\mathcal{J}^{2}_{i}\right)(t,x,0)
\label{eq: delta S due  a to a + da c}
\end{equation}
and
\begin{equation}
\begin{split}
\mathcal{J}^{2}_{i}(t,x,y):=&\,
-
\frac{1}{4\pi}
K^{\ }_{ij}\,
\epsilon^{2\nu\rho}\,
\left(
\partial^{\ }_{\nu}\,
a^{\ }_{j,\rho}
\right)(t,x,y)
\\
&\,
+
\frac{e}{2\pi}\,
Q^{\ }_{i}\,
\epsilon^{2\nu\rho}\,
\left(
\partial^{\ }_{\nu}\,
A^{\ }_{\rho}
\right)(t,x,y)
\\
&\,
+
\frac{s}{2\pi}\,
S^{\ }_{i}\,
\epsilon^{2\nu\rho}\,
\left(
\partial^{\ }_{\nu}\,
B^{\ }_{\rho}
\right)(t,x,y).
\end{split}
\label{eq: delta S due  a to a + da d}
\end{equation}
\end{subequations}

The equations of motion 
\begin{equation}
K^{\ }_{ij}\,
\epsilon^{\mu\nu\rho}\,
\partial^{\ }_{\nu}\,
a^{\ }_{j,\rho}
=
e\,\,
Q^{\ }_{i}\,
\epsilon^{\mu\nu\rho}\,
\partial^{\ }_{\nu}\,
A^{\ }_{\rho}
+
s\,
S^{\ }_{i}\,
\epsilon^{\mu\nu\rho}\,
\partial^{\ }_{\nu}\,
B^{\ }_{\rho}
\end{equation}
for the dynamical gauge field $a$ dictate here that
\begin{equation}
\mathcal{J}^{\mu}_{i}(t,x,y)=
+
\frac{1}{4\pi}
K^{\ }_{ij}\,
\epsilon^{\mu\nu\rho}\,
\partial^{\ }_{\nu}\,
a^{\ }_{j,\rho}
\end{equation}
for $i=1,\cdots,2N$ and $\mu=0,1,2$.
Hence, the $2N$ components of the quasi-particle 3-current
$\mathcal{J}^{\ }_{i}$ obey the continuity equation
$\partial^{\ }_{\mu}\mathcal{J}^{\mu}_{i}=0$
if
$
\left(
\partial^{\ }_{\mu}\,\partial^{\ }_{\nu}
-
\partial^{\ }_{\nu}\,\partial^{\ }_{\mu}
\right)
a^{\ }_{i,\rho}=0
$
holds for any $i=1,\cdots,2N$ and $\rho=0,1,2$.

We now assume that the $2N$-tuplet
$\chi$ is constant along the boundary $\partial\Omega$ for all times,
\begin{equation}
\left(\partial^{\ }_{x}\chi^{\ }_{i}\right)(t,x,y=0)=
\left(\partial^{\ }_{t}\chi^{\ }_{i}\right)(t,x,y=0)=0
\label{eq: restriction a to a+da weak}
\end{equation}
for $i=1,\cdots,2N$.
In this case, each component $\chi^{\ }_i$ 
can be pulled outside the integral in Eq.%
~(\ref{eq: delta S due  a to a + da c})
yielding 
\begin{equation}
\delta\mathcal{S}=
\chi^{\ }_i\,
\int\limits_{-\infty}^{+\infty}\mathrm{d}\,t
\int\limits_{-\infty}^{+\infty}\mathrm{d}\,x\;
\mathcal{J}^{2}_{i}(t,x,0).
\end{equation}
Gauge invariance, i.e., $\delta\mathcal{S}=0$,
is then achieved if, in addition to the restriction%
~(\ref{eq: restriction a to a+da weak}),
we demand that there is no net accumulation of quasi-particle charge 
along the boundary arising from the quasi-particle current normal 
to the boundary, i.e.,
\begin{equation}
0=
\int\limits_{-\infty}^{+\infty}\mathrm{d}\,t
\int\limits_{-\infty}^{+\infty}\mathrm{d}\,x\;
\mathcal{J}^{2}_{i}(t,x,0).
\label{eq: condition for gauge symmetry is chi constant}
\end{equation}
Observe that the stronger condition
\begin{equation}
\chi^{\ }_{i}(t,x,y=0)=0
\label{eq: restriction a to a+da strong}
\end{equation}
for $i=1,\cdots,2N$
achieves gauge invariance, i.e., $\delta\mathcal{S}=0$, without imposing
condition%
~(\ref{eq: condition for gauge symmetry is chi constant}).

Now that we understand under what conditions the quantum field theory
with the action~(\ref{eq:edge-CS})
is gauge invariant with the choice~(\ref{eq: def Omega})
for $\Omega$, we are ready to construct the bulk-edge correspondence. 
To this end, we are going to extract from the dynamical gauge
field $a$ degrees of freedom that are localized on the 
edge $\partial\Omega$ and invariant under the gauge
transformations induced by Eqs.%
~(\ref{eq: delta S due  a to a + da a}),%
~(\ref{eq: restriction a to a+da weak}),
and~(\ref{eq: condition for gauge symmetry is chi constant})
on the edge $\partial\Omega$.

\medskip

\subsection{
Bulk-edge correspondence
        }

We start by fixing the gauge of the $2N$ Abelian Chern-Simons fields 
through the conditions
\begin{subequations}
\begin{equation}
a^{\ }_0=
K^{-1}\,
V\,
a^{\ }_{1}.
\label{eq:Gauge conditions}
\end{equation}
We demand here 
that $V$ is a symmetric, positive definite $2N\times2N$ matrix 
that satisfies
\begin{equation}
V=
\Sigma^{\ }_{1}\,
V\,
\Sigma^{\ }_{1},
\label{eq:V and Sigma1 condition}
\end{equation}
where the $2N\times2N$ matrices
\begin{equation}
\Sigma^{\ }_{\rho}:=
\sigma^{\ }_{\rho}\otimes\openone^{\ }_{N},\qquad \rho=1,2,3
\label{eq: TRS conditions on K and Q c}
\end{equation}
are defined by taking the tensor product between any of 
the Pauli matrices $\sigma^{\ }_{1}$, $\sigma^{\ }_{2}$,
and $\sigma^{\ }_{3}$ and the unit $N\times N$ matrix 
$\openone^{\ }_{N}$.
Condition~\eqref{eq:V and Sigma1 condition} guarantees that
the gauge condition~\eqref{eq:Gauge conditions} 
is consistent with reversal of time defined by
\begin{equation}
\label{TRS on the bulk a fields}
a^{\ }_{\mu}(t,x,y)
\stackrel{\mathcal{T}}{\rightarrow}
-g^{\mu\nu}
\Sigma^{\ }_{1}\,
a^{\ }_{\nu}(-t,x,y).
\end{equation}
\end{subequations}
Indeed,
the gauge condition~\eqref{eq:Gauge conditions} 
then transforms under reversal of time into
\begin{equation}
-
\Sigma^{\ }_{1}\,
a^{0}(-t,x,y)
=
K^{-1}\,
V\,
\Sigma^{\ }_{1}\,
a^{\ }_{1}(-t,x,y),
\end{equation}
which, upon using $K^{-1}=-\Sigma^{\ }_{1}\,K^{-1}\,\Sigma^{\ }_{1}$, coincides with 
Eq.~\eqref{eq:Gauge conditions} if and only if we impose
condition~\eqref{eq:V and Sigma1 condition}.

Next, we use the gauge conditions~\eqref{eq:Gauge conditions}
to eliminate the time components $a^{\ }_0$ 
of the dynamical gauge fields from the theory.
For that, observe that their equations of motion
\begin{subequations}
\label{eq: solving tide f 12 =0}
\begin{equation}
0=
\frac{
\delta \mathcal{S}^{\ }_{K}
     }
     {
\delta a^{\ }_{0}
     }
\Longleftrightarrow
\partial^{\ }_{1}
a^{\ }_{2}
-
\partial^{\ }_{2}
a^{\ }_{1}
=0,
\label{eq: solving tide f 12 =0 a}
\end{equation}
which require the vanishing of their field strengths, 
are automatically satisfied if
\begin{equation}
a^{\ }_{1}=
\partial^{\ }_{1}
\Phi,
\qquad
a^{\ }_{2}=
\partial^{\ }_{2}
\Phi,
\label{eq: solving tide f 12 =0 b}  
\end{equation}
for
\begin{equation}
\left(
\partial^{\ }_{1}
\partial^{\ }_{2}
-
\partial^{\ }_{2}
\partial^{\ }_{1}
\right)
\Phi
=0
\label{eq: solving tide f 12 =0 c}  
\end{equation}
\end{subequations} 
then follows if the $2N$ components
$\Phi^{\ }_{i}$ 
of the vector field $\Phi$
are smooth for $i=1,\cdots,2N$.

\medskip
\begin{widetext}
We rewrite the kinetic part~(\ref{eq:edge-CS b}) 
of the action~(\ref{eq:edge-CS a})
using the gauge conditions%
~\eqref{eq:Gauge conditions}
and the equations of motion%
~\eqref{eq: solving tide f 12 =0 a}
and subsequently substitute the gauge fields $\Phi$ defined in 
Eq.~(\ref{eq: solving tide f 12 =0 b}):
\begin{equation}
\begin{split}
\mathcal{S}^{\ }_{K}
=&\,
-
\frac{\epsilon^{0\nu\lambda}}{4\pi}
\int\limits_{-\infty}^{+\infty}\mathrm{d}t\, 
\int\limits_{-\infty}^{+\infty}\mathrm{d}x\, 
\int\limits_{0}^{+\infty}\mathrm{d}y\, 
\left(
-
a^{\mathsf{T}}_{\nu}\,
K\,
\partial^{\ }_{0}\,
a^{\ }_{\lambda}
+
a^{\mathsf{T}}_{\nu}\,
V\,
\partial^{\ }_{\lambda}\,
a^{\ }_{1}
\right)
\\
=&\,
-
\frac{\epsilon^{0\nu\lambda}}{4\pi}
\int\limits_{-\infty}^{+\infty}\mathrm{d}t\, 
\int\limits_{-\infty}^{+\infty}\mathrm{d}x\, 
\int\limits_{0}^{+\infty}\mathrm{d}y\, 
\left(
\partial^{\ }_\nu\Phi
\right)^{\s{T}}
\left(
K\,
\partial^{\ }_{0}\,
\partial^{\ }_\lambda\Phi
-
V\,
\partial^{\ }_{\lambda}\,
\partial^{\ }_1\Phi
\right)
\\
=
&\,
-
\frac{\epsilon^{0\nu\lambda}}{4\pi}
\int\limits_{-\infty}^{+\infty}\mathrm{d}t\, 
\int\limits_{-\infty}^{+\infty}\mathrm{d}x\, 
\int\limits_{0}^{+\infty}\mathrm{d}y\, 
\partial^{\ }_\nu
\left(
\Phi^{\s{T}}
\,
K\,
\partial^{\ }_{0}\,
\partial^{\ }_\lambda\Phi
-
\Phi^{\s{T}}
\,
V\,
\partial^{\ }_{\lambda}\,
\partial^{\ }_1\Phi
\right)
.
\end{split}
\end{equation}
We shall demand that 
$
\Phi(t,\bm{x})
$
vanishes for $|\bm{x}|\to\infty$,
in which case
\begin{equation}
\begin{split}
\mathcal{S}^{\ }_{K}
=&\,
-
\frac{1}{4\pi}
\int\limits_{-\infty}^{+\infty}\mathrm{d}t\, 
\int\limits_{-\infty}^{+\infty}\mathrm{d}x\, 
\left(
\Phi^{\mathsf{T}}\,
K\,\partial^{\ }_{0}\partial^{\ }_{1}\Phi
-
\Phi^{\mathsf{T}}\,
V\,\partial^{\ }_{1}\,\partial^{\ }_{1}\Phi
\right)
(t,x,0)\\
=&\,
\frac{1}{4\pi}
\int\limits_{-\infty}^{+\infty}\mathrm{d}t\, 
\int\limits_{-\infty}^{+\infty}\mathrm{d}x\, 
\left[
\left(\partial^{\ }_{1}\Phi\right)^{\mathsf{T}}\,
K\,\partial^{\ }_{0}\Phi
-
\left(\partial^{\ }_{1}\Phi\right)^{\mathsf{T}}\,
V\,\partial^{\ }_{1}\Phi
\right]
(t,x,0).
\end{split}
\label{S_KforPhi}
\end{equation}
\end{widetext}
\medskip

Under the gauge transformation%
~(\ref{eq: delta S due  a to a + da a})
subject to the constraints%
~(\ref{eq: restriction a to a+da weak})
and%
~(\ref{eq: condition for gauge symmetry is chi constant})
the $2N$-tuplet $\Phi$ transforms as
\begin{equation}
\Phi(t,\bs{x})\to
\Phi(t,\bs{x})
+
\chi.
\label{eq: trsf law Phi}
\end{equation}
The fact that $\chi$ is independent of time $t$ and space $x$
implies that
(a) the edge theory~(\ref{S_KforPhi}) is unchanged
under Eq.~(\ref{eq: trsf law Phi}), as anticipated,
and (b)
$\left(\partial^{\ }_{1}\Phi\right)(t,x,0)$ 
and 
$\left(\partial^{\ }_{0}\Phi\right)(t,x,0)$ 
are unchanged under Eq.~(\ref{eq: trsf law Phi})
and therefore are physical degrees of freedom at the edge. 
Their dynamics are controlled by the non-universal matrix $V$, 
which is fixed by microscopic details of the physical system
near the edge.

So far, we have discussed only the kinetic part of the action.
Let us now discuss the couplings to the external gauge potentials 
$A$ and $B$ given by the actions~(\ref{eq:edge-CS c}) 
and~(\ref{eq:edge-CS d}), respectively.
We assume that the external gauge field
$A$ is chosen so that (i) all its components are independent of $y$, i.e.,
\begin{subequations}
\label{eq: restricting Amu to boundary}
\begin{equation}
A^{\ }_{\mu}(t,x,y)=
A^{\ }_{\mu}(t,x),
\label{eq: restricting Amu to boundary a}
\end{equation}
for $\mu=0,1,2$
and (ii) they generate the Maxwell equations in
a one-dimensional space defined by the boundary $y=0$, i.e.,
\begin{equation}
A^{\ }_{2}(t,x)=0
\label{eq: restricting Amu to boundary b}
\end{equation}
\end{subequations}
for all times $t$ and for all positions $x$ along the 
one-dimensional boundary $y=0$. 
Using (i) and (ii), we can recast $\mathcal{S}^{\ }_{Q}$ as
\begin{equation}
\begin{split}
\mathcal{S}^{\ }_{Q}=&\,
+
\frac{e}{2\pi}
\int\limits_{\mathbb{R}}\mathrm{d}t\,
\int\limits_{\Omega}\mathrm{d}^{2}\bm{x}\,
Q^{\ }_{i}\,
\epsilon_{\ }^{2\nu\rho}\,
a^{\ }_{i,2}\,
\partial^{\ }_{\nu}\,
A^{\ }_{\rho}
\\
=&\,
+
\frac{e}{2\pi}
\int\limits_{\mathbb{R}}\mathrm{d}t\,
\int\limits_{\Omega}\mathrm{d}^{2}\bm{x}\,
Q^{\ }_{i}\,
\epsilon_{\ }^{2\mu\nu}\,
\partial^{\ }_{2}\,
\left(\Phi^{\ }_{i}\partial^{\ }_{\mu}\,
A^{\ }_{\nu}\right)
\\
=&\,
-
\frac{e}{2\pi}
\int\limits_{-\infty}^{\infty}\mathrm{d}t\,
\int\limits_{-\infty}^{\infty}\mathrm{d}x\,
\left(\epsilon_{\ }^{\mu\nu}\,
A^{\ }_{\mu}\,
Q^{\s{T}}\,
\partial^{\ }_{\nu}\, 
\Phi\right)(t,x,0).
\label{eq:edge-Q derivation}	
\end{split}
\end{equation}
On the last line, the Levi-Civita tensor is defined
for $(1+1)$ space and time.

Furthermore, 
the very same manipulations that lead to
Eq.~(\ref{eq:edge-Q derivation})
can be carried out
on $\mathcal{S}^{\ }_{S}$
to deliver 
\begin{equation}
\begin{split}
\mathcal{S}^{\ }_{S}=&\,
-
\frac{s}{2\pi}\,
\int\limits_{-\infty}^{+\infty}\mathrm{d}t
\int\limits_{-\infty}^{+\infty}\mathrm{d}x
\left(
\epsilon^{\mu\nu}\,
B^{\ }_{\mu}\,
S^{\mathsf{T}}\,
\partial^{\ }_{\nu}\,
\Phi
\right)(t,x,0).
\end{split}
\label{eq: trsf SQ B}
\end{equation}

Finally, the operation of time reversal stated in Eq.%
~\eqref{TRS on the bulk a fields}
in the bulk reduces on the boundary to the transformation law
\begin{equation}
\begin{split}
&
a^{\ }_{1}(t,x)=
\left(\partial^{\ }_{x}\Phi\right)(t,x)
\\
&\qquad\qquad
\stackrel{\mathcal{T}}{\rightarrow}
\Sigma^{\ }_{1}\,
a^{\ }_{1}(-t,x)=
\left(
\partial^{\ }_{x}\,
\Sigma^{\ }_{1}\,
\Phi
\right)(-t,x).
\end{split}
\end{equation}
The transformation law of the $2N$-tuplet
$\Phi$ under reversal of time is thus only fixed
unambiguously up to an additive constant $2N$-tuplet.
The choice 
\begin{subequations}
\begin{equation}
\Phi(t,x)\stackrel{\mathcal{T}}{\rightarrow}
\Sigma^{\ }_{1}\,
\Phi(-t,x)
+
\pi
K^{-1}\,
\Sigma^{\downarrow}\,
Q,
\label{eq: TRS edge theory a}
\end{equation}
with 
\begin{equation}
\Sigma^{\uparrow}:=
\frac{1}{2}
\left(
\Sigma^{\ }_{0}
+
\Sigma^{\ }_{3}
\right),
\qquad
\Sigma^{\downarrow}:=
\frac{1}{2}
\left(
\Sigma^{\ }_{0}
-
\Sigma^{\ }_{3}
\right),
\label{eq: TRS edge theory b}
\end{equation}
\end{subequations}
guarantees that at least one Kramers doublet of fermions
exists as local fields in the edge theory, as was shown in Ref.%
~\onlinecite{Neupert11b}.

\medskip
\subsection{
Fermi-Bose edge correlation functions
           }
\label{subsec: Fermi-Bose edge correlation functions}

Local excitations on the edge can be classified into two groups.
There are quasiparticle excitations that carry rational charges and obey
fractional statistics. There are Fermi-Bose excitations that
carry integer charges and obey Fermi or Bose statistics.
The former excitations are built from vertex operators of the form
\begin{subequations}
\label{eq: def vertx operators}
\begin{equation}
V^{\mathrm{qp}}_{i}(t,x):=
e^{
-\mathrm{i}\Phi^{\ }_{i}(t,x)
  }
\label{eq: def vertx operators a}
\end{equation}
that are labeled by the flavor index $i=1,\cdots,2N$.
The latter excitations are built from the vertex operators of the form
\begin{equation}
V^{\mathrm{fb}}_{i}(t,x):=
e^{
-\mathrm{i} K^{\ }_{ij}\,\Phi^{\ }_{j}(t,x)
  },
\label{eq: def vertx operators b}
\end{equation}
\end{subequations}
that are also labeled by the flavor index $i=1,\cdots,2N$.
Establishing the statistics under exchange obeyed
by these vertex operators can be achieved by computing their
correlation functions, as we now show for the Fermi-Bose 
operators.

We shall choose for $\Omega$ a disk of unit radius centered
at the origin of the complex plane with coordinate $z\in\mathbb{C}$.
Thus, the boundary $\partial\,\Omega$ is the unit circle centered
at the origin of $\mathbb{C}$.
We are after the correlation function
\begin{widetext}
\begin{equation}
\begin{split}
&
\Psi
\left(
\left\{
z^{\ }_{1,1},\bar{z}^{\ }_{1,1},
\cdots,
z^{\ }_{1,n^{\ }_{1}},\bar{z}^{\ }_{1,n^{\ }_{1}}
\right\};
\cdots;
\left\{
z^{\ }_{2N,1},\bar{z}^{\ }_{2N,1},
\cdots,
z^{\ }_{2N,n^{\ }_{2N}},\bar{z}^{\ }_{2N,n^{\ }_{2N}}
\right\}
\right):=
\\
&
\qquad
\qquad
\left\langle
e^{\mathcal{Q}}\,
V^{\mathrm{fb}}_{1}(z^{\ }_{1,1},\bar{z}^{\ }_{1,1})
\times\cdots\times
V^{\mathrm{fb}}_{1}(z^{\ }_{1,n^{\ }_{1}},\bar{z}^{\ }_{1,n^{\ }_{1}})
\times\cdots\times
V^{\mathrm{fb}}_{2N}(z^{\ }_{2N,1},\bar{z}^{\ }_{2N,1})
\times\cdots\times
V^{\mathrm{fb}}_{2N}(z^{\ }_{2N,n^{\ }_{2N}},\bar{z}^{\ }_{2N,n^{\ }_{2N}})
\right\rangle
\end{split}
\label{eq: def FB correlation fct}
\end{equation}
\end{widetext}
where the angular bracket denotes an expectation value
using the quantum field theory with the action%
~(\ref{S_KforPhi})
and $\mathcal{Q}$ is a so-called background charge.
This correlation function fixes the positions of 
$n^{\ }_{i}$ particles of flavor $i=1,\cdots,2N$
at the locations
$z^{\ }_{i,1}$,
$z^{\ }_{i,2}$,
$\cdots$,
$z^{\ }_{i,n^{\ }_{i}}$
along the unit circle. We are omitting any reference to the
time $t$ since all Fermi-Bose vertex operators are taken
at equal time.

We shall use the rules that 
\begin{widetext}
\begin{equation}
\left\langle
\tilde{\Phi}^{\ }_{I}(z,\bar{z})
\tilde{\Phi}^{\ }_{J}(w,\bar{w})
\right\rangle
=
\begin{cases}
\log(z-w),
&
\hbox{if $I=J=1,\cdots,N$,}
\\
\log(\bar{z}-\bar{w}),
&
\hbox{if $I=J=N+1,\cdots,2N$,}
\\
0
&
\hbox{otherwise,}
\end{cases}
\label{eq: def two point functions in I basis}
\end{equation}
\end{widetext}
where the capitalized index $I=1,\cdots,2N$
labels the basis of $\mathbb{R}^{2N}$ for which
the $K$ matrix is represented by the diagonal matrix made of
the signature of its eigenvalues
\begin{equation}
\Sigma^{\ }_{3}=
\left(W^{-1}\right)^{\mathsf{T}}\,
K\,
\left(W^{-1}\right).
\label{eq: def W}
\end{equation}
Observe that the linear transformation
$W$
needs neither be integer-valued nor have unit determinant.
It is a mere useful device to compute
the correlation function%
~(\ref{eq: def FB correlation fct}).
The relationship between the co-ordinates
$\tilde{\Phi}^{\ }_{I}$ and $\Phi^{\ }_{j}$ is linear and given by
\begin{equation}
\tilde{\Phi}^{\ }_{I}=
W^{\ }_{Ij}\,
\Phi^{\ }_{j},
\qquad
I=1,\cdots,2N,
\end{equation}
where the summation convention for the repeated small case indices
is used. 

In order to take advantage of
Eq.~(\ref{eq: def two point functions in I basis})
when evaluating
Eq.~(\ref{eq: def FB correlation fct}),
we use the decomposition
\begin{widetext}
\begin{subequations}
\label{eq: projective decomposition}
\begin{equation}
\begin{split}
K^{\ }_{ij} \Phi^{\ }_{j}(z,\bar{z})=&\,
W^{\ }_{Ii}\,\left(\Sigma^{\ }_{3}\right)^{\ }_{IJ}\,
W^{\ }_{Jj}
\,
\Phi^{\ }_{j}(z,\bar{z})
\\
=&\,
W^{\ }_{Ii}\,
\left(
\Sigma^{\uparrow}_{IJ}
-
\Sigma^{\downarrow}_{IJ}
\right)\,
\tilde{\Phi}^{\ }_{J}(z,\bar{z})
\\
=&\,
\left(
W^{\ }_{Ii}\,
\Sigma^{\uparrow}_{IJ}\,
\tilde{\Phi}^{\ }_{J}\right)
(z)
-
\left(
W^{\ }_{Ii}\,
\Sigma^{\downarrow}_{IJ}
\tilde{\Phi}^{\ }_{J}\right)
(\bar{z})
\end{split}
\label{eq: projective decomposition a}
\end{equation}
for $i=1,\cdots,2N$
where the matrices 
$\Sigma^{\uparrow}$ 
and
$\Sigma^{\downarrow}$
were defined in Eq.~\eqref{eq: TRS edge theory b}.
Under the decomposition%
~(\ref{eq: projective decomposition a}),
any Fermi-Bose vertex operator%
~(\ref{eq: def vertx operators b})
occurring in the correlation function%
~(\ref{eq: def FB correlation fct})
becomes
\begin{equation}
\begin{split}
V^{\mathrm{fb}}_{i}(z,\bar{z})=&\,
e^{
-\mathrm{i} K^{\ }_{ij}\,\Phi^{\ }_{j}(z,\bar{z})
  }
=
\exp
\left(
-\mathrm{i}
\left( 
W^{\ }_{Ii}\,
\Sigma^{\uparrow}_{IJ}\,
\tilde{\Phi}^{\ }_{J}
\right)(z)
\right)
\times
\exp
\left(
+\mathrm{i}
\left( 
W^{\ }_{Ii}\,
\Sigma^{\downarrow}_{IJ}\,
\tilde{\Phi}^{\ }_{J}
\right)
(\bar{z})
\right).
\end{split}
\label{eq: projective decomposition b}
\end{equation}
We shall also decompose accordingly the background charge 
$\mathcal{Q}=\mathcal{Q}^{\ }_{\uparrow}+\mathcal{Q}^{\ }_{\downarrow}$.
\end{subequations}
Now,
\begin{equation}
\begin{split}
\Psi
\left(\cdots;
z^{\ }_{i,1},\cdots,\bar{z}^{\ }_{i,n^{\ }_{i}};\cdots
\right)=&\,
\exp
\left(
\mathcal{Q}^{\ }_{\uparrow}
+
\frac{1}{2}
\left\langle
\left(
+
\sum\limits_{i=1}^{2N}
\sum\limits_{a^{\ }_{i}=1}^{n^{\ }_{i}}
\left( 
W^{\ }_{Ii}\,
\Sigma^{\uparrow}_{IJ}\,
\tilde{\Phi}^{\ }_{J}
\right)
(z^{\ }_{i,a^{\ }_{i}})
\right)^{2}
\right\rangle
\right)
\\
&\,
\times
\exp
\left(
\mathcal{Q}^{\ }_{\downarrow}
+
\frac{1}{2}
\left\langle
\left(
-
\sum\limits_{i=1}^{2N}
\sum\limits_{a^{\ }_{i}=1}^{n^{\ }_{i}}
\left( 
W^{\ }_{Ii}\,
\Sigma^{\downarrow}_{IJ}\,
\tilde{\Phi}^{\ }_{J}
\right)
(\bar{z}^{\ }_{i,a^{\ }_{i}})
\right)^{2}
\right\rangle
\right)
\\
=&\,
\left[
\prod_{i=1}^{2N}
\prod_{1\leq a^{\ }_{i}<b^{\ }_{i}\leq n^{\ }_{i}}
\left(
z^{\ }_{i,a^{\ }_{i}}
-
z^{\ }_{i,b^{\ }_{i}}
\right)^{
W^{\ }_{Ii}
\Sigma^{\uparrow}_{IJ}
W^{\ }_{Ji}
        }
\left(
\bar{z}^{\ }_{i,a^{\ }_{i}}
-
\bar{z}^{\ }_{i,b^{\ }_{i}}
\right)^{
W^{\ }_{Ii}
\Sigma^{\downarrow}_{IJ}
W^{\ }_{Ji}
        }
\right]
\\
&\times\,
\left[
\prod_{1\leq i<j\leq 2N}
\prod_{a^{\ }_{i},b^{\ }_{j}=1}^{n^{\ }_{i}}
\left(
z^{\ }_{i,a^{\ }_{i}}
-
z^{\ }_{j,b^{\ }_{j}}
\right)^{
W^{\ }_{Ii}
\Sigma^{\uparrow}_{IJ}
W^{\ }_{Jj}
        }
\left(
\bar{z}^{\ }_{i,a^{\ }_{i}}
-
\bar{z}^{\ }_{j,b^{\ }_{j}}
\right)^{
W^{\ }_{Ii}
\Sigma^{\downarrow}_{IJ}
W^{\ }_{Jj}
        }
\right].
\end{split}
\label{eq: final result for correlation fct}
\end{equation}
\end{widetext}
The role of the background charge
is to guarantee ``charge neutrality''. 
Observe that for any pair $i,j=1,\cdots,2N$, the exponents
\begin{equation}
\alpha^{\ }_{ij}:=
W^{\ }_{Ii}\,
\Sigma^{\uparrow}_{IJ}\,
W^{\ }_{Jj}
\end{equation}
and
\begin{equation}
\beta^{\ }_{ij}:=
W^{\ }_{Ii}\,
\Sigma^{\downarrow}_{IJ}\,
W^{\ }_{Jj}
\end{equation}
can be an irrational number!
Nevertheless if 
$z^{\ }_{i,a^{\ }_{i}}\neq z^{\ }_{j,b^{\ }_{j}}$
the correlation function%
~(\ref{eq: final result for correlation fct}) 
is single-valued, 
for it is the product of functions of the form
\begin{subequations}
\begin{widetext}
\begin{equation}
\begin{split}
f(z^{\ }_{i,a^{\ }_{i}},z^{\ }_{j,b^{\ }_{j}}):=&\,
\left(
z^{\ }_{i,a^{\ }_{i}}
-
z^{\ }_{j,b^{\ }_{j}}
\right)^{\alpha^{\ }_{ij}}
\left(
\bar{z}^{\ }_{i,a^{\ }_{i}}
-
\bar{z}^{\ }_{j,b^{\ }_{j}}
\right)^{\beta^{\ }_{ij}}
\\
=&\,
\left(
z^{\ }_{i,a^{\ }_{i}}
-
z^{\ }_{j,b^{\ }_{j}}
\right)^{\alpha^{\ }_{ij}-\beta^{\ }_{ij}+\beta^{\ }_{ij}}
\left(
\bar{z}^{\ }_{i,a^{\ }_{i}}
-
\bar{z}^{\ }_{j,b^{\ }_{j}}
\right)^{\beta^{\ }_{ij}}
\\
=&\,
\left(
z^{\ }_{i,a^{\ }_{i}}
-
z^{\ }_{j,b^{\ }_{j}}
\right)^{K^{\ }_{ij}}
\left|
z^{\ }_{i,a^{\ }_{i}}
-
z^{\ }_{j,b^{\ }_{j}}
\right|^{2\beta^{\ }_{ij}}
\end{split}
\label{eq: def f(z bar z)}
\end{equation}
\end{widetext}
where one verifies that
\begin{equation}
\alpha^{\ }_{ij}
-
\beta^{\ }_{ij}=
K^{\ }_{ij}
\end{equation}
\end{subequations}
is integer valued. This is consistent with the fact that
the vertex operators%
~(\ref{eq: def vertx operators b})
describe either Fermions or Bosons.

\medskip
\section{
Wave functions
        }
\label{sec: Wave functions}

The family of topological quantum field theories defined by
Eqs.%
~(\ref{eq:intro-CS-BF}) and%
~(\ref{eq: def equivalence classes})
encode the universal properties of a family of 
time-reversal symmetric fractional quantum liquids.

Any connection to a microscopic realization of a
time-reversal symmetric fractional quantum liquid
whose universal properties are captured by
Eqs.%
~(\ref{eq:intro-CS-BF}) and%
~(\ref{eq: def equivalence classes})
must, however, be supplied. 

For example, the very definition of an electron operator
is ambiguous for any equivalence class of
topological quantum field theories defined by
Eqs.%
~(\ref{eq:intro-CS-BF}) and%
~(\ref{eq: def equivalence classes}).
First, there is no unique definition of a local fermion in any
topological quantum field theory of the form
Eqs.%
~(\ref{eq:intro-CS-BF}) and%
~(\ref{eq: def equivalence classes})
that admits a hierarchical representation with one type
of Kramers degenerate pairs of fermions,
for this representation is equivalent to 
the symmetric representation that admits $N$
distinct types of Kramers degenerate pair of fermions
[see Eq.~(\ref{eq: def trsf from hierarchy to symmetric basis})].
Second, for any representation of the universal data $(K,Q,S)$ 
from Eq.~(\ref{eq:intro-K-matrix})
that admits fermions, a basis set of functions
must be supplied to construct a representation of the microscopic
electron operator. This basis set of functions is usually 
provided by some reference single-particle electron basis.

In the context of the FQHE
observed in a GaAs accumulation layer,
the basis set of functions is the single-particle basis
of the Landau Hamiltonian describing an electron moving in a 
plane perpendicular to a uniform magnetic field.
However, the Landau basis set and, in particular, the basis set
for the lowest Landau level, is not appropriate for the recently 
discovered fractional quantum Hall phases in lattice models
without external magnetic field.%
~\cite{Neupert11a,Sheng11,Wang11a,Regnault11}

With this caveat in mind, we are going
to construct some wave functions
using the data $(K,Q,S)$ 
from Eq.~(\ref{eq:intro-K-matrix})
as the universal input
and using the Landau wave functions spanning the lowest Landau level
as the microscopic input. We do this out of simplicity in view
of the elegant analytic properties of these single-particle functions.
Hence, we choose the symmetric gauge for which
the Slater determinant in the lowest Landau level is
\begin{widetext}
\begin{subequations}
\begin{equation}
\begin{split}
\Psi^{\ }_{\nu^{\ }_{\mathsf{i}}=1}
\left(
\left\{z^{\ }_{\mathsf{i}},\bar{z}^{\ }_{\mathsf{i}}\right\}^{\ }_{n^{\ }_{\mathsf{i}}}
\right):=&\,
\left[
\prod_{1\leq k<l\leq n^{\ }_{\mathsf{i}}}
\left(
z^{\ }_{\mathsf{i},k}
-
z^{\ }_{\mathsf{i},l}
\right)
\right]
\times
\prod_{k=1}^{n^{\ }_{\mathsf{i}}}
\exp
\left(
-
\frac{
\bar{z}^{\ }_{\mathsf{i},k}
z^{\ }_{\mathsf{i},k}
     }
     {
4\ell^{2}
     }
\right)
\end{split}
\end{equation}
for $n^{\ }_{\mathsf{i}}$ 
electrons  labeled by the flavor index $\mathsf{i}=1,\cdots,N$
while it is 
\begin{equation}
\begin{split}
\Psi^{\ }_{\nu^{\ }_{\mathsf{i}}=1}
\left(
\left\{w^{\ }_{\mathsf{i}},\bar{w}^{\ }_{\mathsf{i}}\right\}^{\ }_{n^{\ }_{\mathsf{i}}}
\right):=&\,
\left[
\prod_{1\leq k<l\leq n^{\ }_{\mathsf{i}}}
\left(
\bar{w}^{\ }_{\mathsf{i},k}
-
\bar{w}^{\ }_{\mathsf{i},l}
\right)
\right]
\times
\prod_{k=1}^{n^{\ }_{\mathsf{i}}}
\exp
\left(
-
\frac{
\bar{w}^{\ }_{\mathsf{i},k}
w^{\ }_{\mathsf{i},k}
     }
     {
4\ell^{2}
     }
\right)
\end{split}
\end{equation}
\end{subequations}
\end{widetext}
for $n^{\ }_{N+\mathsf{i}}$  electrons
labeled by the flavor index $N+\mathsf{i}=N+1,\cdots,2N$.
Here, 
\begin{equation}
\left\{z^{\ }_{\mathsf{i}},\bar{z}^{\ }_{\mathsf{i}}\right\}^{\ }_{n^{\ }_{\mathsf{i}}}
:=
\left\{
z^{\ }_{\mathsf{i},1},\cdots,z^{\ }_{\mathsf{i},n_{\s{i}}},
\bar{z}^{\ }_{\mathsf{i},1},\cdots,\bar{z}^{\ }_{\mathsf{i},n_{\s{i}}}
\right\}
\end{equation}
denotes the complex coordinates of the particles of the first $N$ flavors, with 
$\bar{z}$ denoting their complex conjugates, and likewise 
$\left\{w^{\ }_{\mathsf{i}},\bar{w}^{\ }_{\mathsf{i}}\right\}^{\ }_{n^{\ }_{\mathsf{i}}}$
denotes the complex coordinates for the last $N$ flavors;
$\ell^{2}:=\phi^{\ }_{0}/(2\pi|B|)$ 
is the square of the magnetic length $\ell^{\ }$
in the presence of the uniform
magnetic field of magnitude $|B|$, 
$\phi^{\ }_{0}:=2\,\pi/e$ 
is the quantum of magnetic flux,
$|\Omega||B|$ is the magnitude of the flux threading the disk $\Omega$,
and
$\nu^{\ }_{\mathsf{i}}:=(n^{\ }_{\mathsf{i}}\phi^{\ }_{0})/(|\Omega||B|)$
represents the filling fraction of the lowest Landau level.

It remains to decide on the number of electron flavors,
a microscopic input. 
We shall assume that the hierarchical (symmetric)
representation corresponds to a single pair ($N$ pairs) of microscopic flavors
of electrons forming a Kramers doublet ($N$ Kramers doublets). 
We begin with the wave function for $N=1$, 
in which case there is no distinction between the two representations.
We then work out examples with 
$N=2$ in the symmetric and the hierarchical representation.

\subsection{
Wave function for $N=1$
           }

We choose the universal data to be
\begin{equation}
K=
\begin{pmatrix}
+m
&
0
\\
0
&
-m
\end{pmatrix}
\in\mathrm{GL}(2,\mathbb{Z}),
\qquad
Q=
\begin{pmatrix}
1
\\
1
\end{pmatrix}
\in\mathbb{Z}^{2},
\end{equation} 
for some given positive odd integer $m$.
The spin filling fraction defined in Eq.~(\ref{eq: QSHE})
is
\begin{equation}
\nu^{\ }_{\mathrm{s}}=
\frac{1}{m}.
\end{equation}
The putative ground state wave function 
that generalizes the $\nu=1/m$ single-layer
wave function from Laughlin (see Ref.%
~\onlinecite{Laughlin83a})
to the time-reversal symmetric case is
\begin{widetext}
\begin{equation}
\Psi^{\ }_{1/m}
\left(\left\{z,\bar{z} \right\}^{\ }_n | \left\{w,\bar{w} \right\}^{\ }_n \right)=
\left[
\prod_{
i=1
      }^{n}
\prod_{
j=i+1
      }^{n}
\left(
z^{\ }_{i}
-
z^{\ }_{j}
\right)^{m}
\left(
\bar{w}^{\ }_{i}
-
\bar{w}^{\ }_{j}
\right)^{m}
\right]\times
\prod_{i=1}^{n}
\mathrm{exp}
\left(
-\frac{
\left|z^{\ }_{i}\right|^{2}
+\left|\bar{w}^{\ }_{i}\right|^{2}
}{4\ell^{2}}
\right)
\!
.
\end{equation}
\end{widetext}
By construction, it is invariant under the operation of
time reversal represented by 
\begin{equation}
z^{\ }_{i}
\stackrel{\mathcal{T}}{\rightarrow}
\bar{w}^{\ }_{i},
\qquad
w^{\ }_{i}
\stackrel{\mathcal{T}}{\rightarrow}
\bar{z}^{\ }_{i},
\qquad
i=1,\cdots,n.
\end{equation}
It thus realizes a time-reversal symmetric
fractional incompressible state.
Observe that this wave function factorizes into 
an holomorphic and an antiholomorphic sector.
Time-reversal symmetry forbids a coupling between
the holomorphic and antiholomorphic sector
when $N=1$.

\subsection{
Wave functions in the symmetric representation
           }

We choose the universal data to be
\begin{subequations}
\begin{equation}
K=
\begin{pmatrix}
+\kappa
&
+\Delta
\\
-\Delta
&
-\kappa
\end{pmatrix}
\in\mathrm{GL}(4,\mathbb{Z}),
\qquad
Q=
\begin{pmatrix}
\varrho
\\
\varrho
\end{pmatrix}
\in\mathbb{Z}^{4}.
\end{equation} 
The $2\times2$ matrix $\kappa$ is given by
\begin{equation}
\kappa=
\begin{pmatrix}
m^{\ }_{\mathsf{1}}
&
n
\\
n
&
m^{\ }_{\mathsf{2}}
\end{pmatrix}
\in\mathrm{GL}(2,\mathbb{Z}).
\end{equation}
We impose that the integers 
$m^{\ }_{\mathsf{1}}$
and 
$m^{\ }_{\mathsf{2}}$
are odd and positive 
while the integer $n$ is positive 
[$n\geq0$ is not restrictive in view of Eq.%
~(\ref{eq: example 2 for W})] 
whereby
\begin{equation}
m^{\ }_{\mathsf{1}}m^{\ }_{\mathsf{2}}-n^{2}>0,
\end{equation}
in order for $\kappa$ to be maximally chiral.
In turn, 
\begin{equation}
\Delta=
\begin{pmatrix}
0
&
+d
\\
-d
&
0
\end{pmatrix}\;,
\end{equation}
where the integer
$d\geq 0$
is chosen to be non-negative.
Finally, the charge vector
\begin{equation}
\varrho=
\begin{pmatrix}
1
\\
1
\end{pmatrix}
\end{equation}
\end{subequations}
\\
\noindent
enforces the presence of 4 fermions related pairwise by reversal of time.
We assume that the fermions with the charge vector $Q$
in the topological quantum field theory represent
$2N$ distinct flavors of electrons in a microscopic theory.
For example, each flavor of electrons
could be constrained to move with the dynamics dictated by the
single-particle Landau Hamiltonian
in its own two-dimensional layer
in the presence of a uniform magnetic field pointing up
for the first $N$ layers and down for the next $N$ layers.
If the lowest Landau level of each layer is partially filled,
interactions might select an incompressible ground state.
The spin filling fraction defined in Eq.~(\ref{eq: QSHE})
is
\begin{equation}
\nu^{\ }_{\mathrm{s}}=
\frac{
m^{\ }_{\s{1}}
+
m^{\ }_{\s{2}}
-
2n
     }
     {
m^{\ }_{\s{1}}\,m^{\ }_{\s{2}}
-
n^{2}
+
d^{2}
     }.
\end{equation}
The putative ground state wave function 
that generalizes the $(m^{\ }_{\mathsf{1}},m^{\ }_{\mathsf{2}},n)$
bilayer wave function from Halperin (see Ref.~\onlinecite{Halperin83})
to the time-reversal symmetric case is
\begin{widetext}
\begin{equation}
\begin{split}
&
\Psi^{\mathrm{symm}}_{m^{\ }_{\mathsf{1}},m^{\ }_{\mathsf{2}},n,d}
\left(
\left\{z^{\ }_{\mathsf{1}},\bar{z}^{\ }_{\mathsf{1}} \right\}
^{\ }_{n^{\ }_{\s{1}}};
\left\{z^{\ }_{\mathsf{2}},\bar{z}^{\ }_{\mathsf{2}} \right\}
^{\ }_{n^{\ }_{\s{2}}}|
\left\{w^{\ }_{\mathsf{1}},\bar{w}^{\ }_{\mathsf{1}} \right\}
^{\ }_{n^{\ }_{\s{1}}};
\left\{w^{\ }_{\mathsf{2}},\bar{w}^{\ }_{\mathsf{2}} \right\}
^{\ }_{n^{\ }_{\s{2}}}
\right)=
\\
&
\qquad
\hphantom{\times\ }
\Psi^{\ }_{1/m^{\ }_{\mathsf{1}}}
\left(
\left\{z^{\ }_{\mathsf{1}},\bar{z}^{\ }_{\mathsf{1}} \right\}
^{\ }_{n^{\ }_{\s{1}}} 
| 
\left\{w^{\ }_{\mathsf{1}},\bar{w}^{\ }_{\mathsf{1}} \right\}
^{\ }_{n^{\ }_{\s{1}}}
\right)
\times
\Psi^{\ }_{1/m^{\ }_{\mathsf{2}}}
\left(
\left\{z^{\ }_{\mathsf{2}},\bar{z}^{\ }_{\mathsf{2}} \right\}
^{\ }_{n^{\ }_{\s{2}}} 
| 
\left\{w^{\ }_{\mathsf{2}},\bar{w}^{\ }_{\mathsf{2}} \right\}
^{\ }_{n^{\ }_{\s{2}}}
\right)
\\
&
\qquad\times
\prod_{
i=1
      }^{n^{\ }_{\mathsf{1}}}
\prod_{
j=1
      }^{n^{\ }_{\mathsf{2}}}
\left(
z^{\ }_{\mathsf{1},i}
-
z^{\ }_{\mathsf{2},j}
\right)^{n}
\left(
\bar{w}^{\ }_{\mathsf{1},i}
-
\bar{w}^{\ }_{\mathsf{2},j}
\right)^{n}
\left(
z^{\ }_{\mathsf{1},i}
-
w^{\ }_{\mathsf{2},j}
\right)^{d}
\left(
\bar{w}^{\ }_{\mathsf{1},i}
-
\bar{z}^{\ }_{\mathsf{2},j}
\right)^{d}
.
\end{split}
\end{equation}
\end{widetext}
By construction, it is invariant under the operation of
time reversal represented by 
\begin{equation}
z^{\ }_{\mathsf{i},i^{\ }_{\mathsf{i}}}
\stackrel{\mathcal{T}}{\rightarrow}
\bar{w}^{\ }_{\mathsf{i},i^{\ }_{\mathsf{i}}},
\qquad
w^{\ }_{\mathsf{i},i^{\ }_{\mathsf{i}}}
\stackrel{\mathcal{T}}{\rightarrow}
\bar{z}^{\ }_{\mathsf{i},i^{\ }_{\mathsf{i}}},
\qquad
i^{\ }_{\mathsf{i}}=1,\cdots,n^{\ }_{\mathsf{i}},
\end{equation}
for $\mathsf{i}=1,2$. It thus realizes a time-reversal symmetric
fractional incompressible state.

\subsection{
Wave functions in the hierarchical representation
           }

We choose the universal data to be
\begin{subequations}
\begin{equation}
K=
\begin{pmatrix}
+\kappa
&
+\Delta
\\
-\Delta
&
-\kappa
\end{pmatrix}
\in\mathrm{GL}(4,\mathbb{Z}),
\qquad
Q=
\begin{pmatrix}
\varrho
\\
\varrho
\end{pmatrix}
\in\mathbb{Z}^{4}.
\end{equation} 
The $2\times2$ matrix $\kappa$ is given by
\begin{equation}
\kappa=
\begin{pmatrix}
+m
&
+1
\\
+1
&
-p
\end{pmatrix}
\in\mathrm{GL}(2,\mathbb{Z})
\end{equation}
where $m$ is a positive odd integer and $p$ is an even
integer larger than zero.
The $2\times2$ matrix $\Delta$ is given by
\begin{equation}
\Delta=
\begin{pmatrix}
0
&
+d
\\
-d
&
0
\end{pmatrix}
\end{equation}
with $d$ any positive integer.
Finally, the charge vector
\begin{equation}
\varrho=
\begin{pmatrix}
1
\\
0
\end{pmatrix}
\end{equation}
\end{subequations}
\\
\noindent
enforces the presence of 2 fermions related pairwise by reversal of time.
The spin filling fraction defined in Eq.~(\ref{eq: QSHE})
is
\begin{equation}
\nu^{\ }_{\mathrm{s}}=
\frac{p}{mp+1-d^{2}}.
\end{equation}
The putative ground state wave function 
that generalizes the $\nu=\frac{p}{mp+1}$ single-layer
wave function from Halperin (see Ref.~\onlinecite{Halperin83})
to the time-reversal symmetric case is
\begin{widetext}
\begin{equation}
\begin{split}
\Psi^{\mathrm{hier}}_{m,-p,1,d}
\left(
\left\{z,\bar{z}\right\}^{\ }_{pn}
|
\left\{w,\bar{w}\right\}^{\ }_{pn}
\right)=
&\,
\hphantom{\times}
\left[
\prod_{i=1}^{n}
\int\limits_{\Omega}
\mathrm{d}^{2}\,
\eta^{\ }_{i}
\int\limits_{\Omega}
\mathrm{d}^{2}\,
\xi^{\ }_{i}
\right]
\times
\Psi^{\ }_{1/m}
\Bigl(
\left\{z,\bar{z}\right\}^{\ }_{pn} 
| 
\left\{w,\bar{w}\right\}^{\ }_{pn}
\Bigr)
\times
\Psi^{\ }_{1/p}
\Bigl(
\left\{\xi,\bar{\xi}\right\}^{\ }_{n} 
| 
\left\{\eta,\bar{\eta}\right\}^{\ }_{n}
\Bigr)
\\
&
\times
\prod_{
i=1
      }^{pn}
\prod_{
j=1
      }^{n}
\left(
z^{\ }_{i}
-
\eta^{\ }_{j}
\right)
\left(
\bar{w}^{\ }_{i}
-
\bar{\xi}^{\ }_{j}
\right)
\left(
z^{\ }_{i}
-
\xi^{\ }_{j}
\right)^{d}
\left(
\bar{w}^{\ }_{i}
-
\bar{\eta}^{\ }_{j}
\right)^{d}
.
\end{split}
\end{equation}
\end{widetext}
By construction, it is invariant under the operation of
time reversal represented by
\begin{subequations} 
\begin{equation}
z^{\ }_{i}
\stackrel{\mathcal{T}}{\rightarrow}
\bar{w}^{\ }_{i},
\qquad
w^{\ }_{i}
\stackrel{\mathcal{T}}{\rightarrow}
\bar{z}^{\ }_{i},
\end{equation}
for $i=1,\cdots,pn$ 
and
\begin{equation}
\xi^{\ }_{i}
\stackrel{\mathcal{T}}{\rightarrow}
\bar{\eta}^{\ }_{i},
\qquad
\eta^{\ }_{i}
\stackrel{\mathcal{T}}{\rightarrow}
\bar{\xi}^{\ }_{i},
\end{equation}
\end{subequations}
for $i=1,\cdots,n$. 
It thus realizes a time-reversal symmetric
fractional incompressible state.

\medskip
\section{
Summary
        }
\label{sec: Summary}

In this paper, we first derived a hierarchy of FQSHEs,
the universal properties of which are encoded by 
equivalence classes of BF theories
of the form
\begin{equation}
\begin{split}
\mathcal{L}=&\,
-
\frac{1}{\pi}
\epsilon^{\mu\nu\lambda}\,
a^{(+)\mathsf{T}}_{\mu}\,
\varkappa\,
\partial^{\ }_{\nu}\,
a^{(-)}_{\lambda}
\\
&\,
+
\frac{e}{\pi}
\epsilon^{\mu\nu\lambda}\,
A^{\ }_{\mu}\,
\varrho^{\mathsf{T}}\,
\partial^{\ }_{\nu}\,
a^{(+)}_{\lambda}
+
\frac{s}{\pi}
\epsilon^{\mu\nu\lambda}\,
B^{\ }_{\mu}\,
\varrho^{\mathsf{T}}\,
\partial^{\ }_{\nu}\,
a^{(-)}_{\lambda}.
\end{split}
\label{eq: summary def theory a}
\end{equation}
The $N\times N$ invertible and integer-valued matrix 
$\varkappa$ couples the
$N$ flavors of the dynamical gauge field $a^{(+)}$ to the
$N$ flavors of the dynamical gauge field $a^{(-)}$.
The $N$-tuplets $a^{(+)}$ and $a^{(-)}$ also couple linearly to the external
gauge fields $A$ and $B$, respectively, through the  
vector $\varrho\in\mathbb{Z}^{N}$, where
the integer $\varrho^{\ }_{\mathsf{i}}$ 
shares the same parity as the integer
$\varkappa^{\ }_{\mathsf{i}\mathsf{i}}$
for $\mathsf{i}=1,\cdots,N$.
Correspondingly, there exists two independent conserved currents,
a charge current associated to the gauge field $a^{(+)}$
and a spin current associated to the gauge field $a^{(-)}$.

Time-reversal symmetry implies the
vanishing of the charge Hall conductivity
\begin{equation}
\sigma^{\ }_{\mathrm{H}}=
\frac{e^{2}}{2\pi}\times \nu=0.
\end{equation}
The non-vanishing spin filling fraction
\begin{subequations}
\begin{equation}
\nu^{\ }_{\mathrm{s}}:=
\varrho^{\s{T}}\,\varkappa^{-1}\,\varrho
\label{eq: spin filling fraction summary}
\end{equation}
can be interpreted as the spin Hall conductance
\begin{equation}
\sigma^{\ }_{\mathrm{sH}}:=
\frac{e}{2\pi}\times\nu^{\ }_{\mathrm{s}}
\end{equation}
\end{subequations}
if the U(1) conservation law associated to the
current of $a^{(-)}$ arises microscopically from
a residual spin-1/2 U(1) (easy plane $XY$)
symmetry. The topological ground state degeneracy, 
if two-dimensional space $\Omega$ has a toroidal geometry,
\begin{equation}
(\mathrm{det}\,\varkappa)^{2}
\end{equation}
is always the square of an integer as a consequence 
of time-reversal symmetry. Equivalent pairs
$(\varkappa, \varrho)$ and $(\varkappa', \varrho')$,
as defined by Eq.~(\ref{eq: def equivalence classes}),
share the same spin Hall conductivity and topological degeneracy.

The theory%
~(\ref{eq: summary def theory a}) 
is topological when two-dimensional space $\Omega$ 
has no boundary, i.e., the Hamiltonian density
associated to the Lagrangian density%
~(\ref{eq: summary def theory a})
vanishes. This is not true anymore if the boundary
$\partial\Omega$ is a one-dimensional manifold.
We have shown that imposing gauge invariance
delivers a gapless theory with all excitations propagating
along the boundary $\partial\Omega$. These excitations can
all be constructed out of $N$ pairs of counter-propagating
chiral bosons whose non-universal velocities 
along the boundary $\partial\Omega$
derive from a gauge-fixing condition.
The stability of this edge theory to the 
(time-reversal symmetric) breaking of
translation invariance along the boundary 
(including the breaking of the spin conservation law
associated to the spin vector $S$)
was studied in Ref.~\onlinecite{Neupert11b}.
The correlation functions for the Fermi-Bose excitations
along the edge were computed and shown to be a product over
the functions~(\ref{eq: def f(z bar z)}).

Finally, we have proposed a time-reversal symmetric 
counterpart to the hierarchy of wave functions that have been proposed
in the context of the FQHE by way of few examples,
the 
$\nu^{\ }_{\mathrm{s}}=1/m$,
$\nu^{\ }_{\mathrm{s}}=p/(mp+1-d^{2})$, 
and $\nu^{\ }_{\mathrm{s}}=
(m^{\ }_{1}+m^{\ }_{2}-2n)/(m^{\ }_{1}m^{\ }_{2}-n^{2}+d^{2})$
sequences.

\medskip
\section*{Acknowledgments}

We gratefully acknowledge useful discussions with 
Maurizio Storni.
This work was supported in part by DOE Grant DEFG02-06ER46316
and by the Swiss National Science Foundation.


\begin{thebibliography}{99}
\bibitem{BFnote}
This terminology has no deep meaning. It arose historically
in the context of topological field theories and refers to
the symbols $b$ and $f$ used to represent a $p$-form $b$ and the curvature
$f$ of a $(d-p)$-form in $(d+1)$-dimensional space and time.%
~\cite{Blau91}
\end{thebibliography}
\end{document}